\pdfoutput=1

\documentclass[journal]{IEEEtran}

\usepackage{cite}
\usepackage[pdftex]{graphicx}
\usepackage[cmex10]{amsmath}
\usepackage{array}
\usepackage{mdwmath}
\usepackage{mdwtab}
\usepackage{flushenda}

\newcommand{\ds}{\displaystyle}
\newcommand{\Ref}[1]{(\ref{#1})}

\begin{document}
\bstctlcite{IEEEexample:BSTcontrol}

\title{Development of 2D Bi-SQUID Arrays \\ with High Linearity}

\author{S.~Berggren,~\IEEEmembership{Member,~IEEE,}
	G.~Prokopenko, 
        P.~Longhini,~\IEEEmembership{Member,~IEEE,}
        A.~Palacios,
        O.~A.~Mukhanov,~\IEEEmembership{Fellow,~IEEE,}
        A.~Leese de Escobar,~\IEEEmembership{Member,~IEEE,}
        B.~J.~Taylor,
        M.~C.~de~Andrade,~\IEEEmembership{Member,~IEEE,}
        M.~Nisenoff,~\IEEEmembership{Life Fellow,~IEEE,}
        R.~L.~Fagaly,~\IEEEmembership{Fellow, ~IEEE,}
        T.~Wong,
        E.~Cho,
        E.~Wong,
        V.~In
\thanks{Manuscript received October 9, 2012. This work is supported in part by the Tactical SIGINT Technology Program N66001-08-D-0154, SPAWAR SBIR contracts N00039-08-C-0024, N66001-09-R-0073, ONR (Code 30), the SPAWAR internal research funding (S\&T) program and HYPRES IR\&D Program.}
\thanks{S. Berggren and A. Palacios are with San Diego State University, San Diego, CA 92182 USA (e-mail: susan\_berggren@yahoo.com).}
\thanks{G. Prokopenko and O.~A. Mukhanov are with HYPRES Inc., Elmsford, NY 10523 USA (phone:	914-592-1190;	fax:	914-347-2239;	e-mail: georgy@hypres.com, mukhanov@hypres.com).}
\thanks{P. Longhini, A. Leese de Escobar, B.~J. Taylor, M.~C. de Andrade, E. Wong and V. In are with SPAWAR SSC Pacific, San Diego, CA 92152 USA.}
\thanks{M. Nisenoff is with M. Nisenoff Associates, Minneapolis, MN 55403 USA.}
\thanks{R.~L. Fagaly is with Quasar Federal Systems, San Diego, CA 92121 USA.}
\thanks{E.~Cho and T.~Wong are with the University of California, San Diego, La Jolla, CA 92093}}

\markboth{1EB-01}
{Berggren \MakeLowercase{\textit{et al.}}: Development of 2D bi-SQUID arrays with high linearity}

\maketitle

\begin{abstract}
We develop a two-dimensional (2D) Superconducting Quantum Interference Filter (SQIF) array based on recently introduced high-linearity tri-junction bi-SQUIDs.  Our bi-SQUID SQIF array design is based on a tight integration of individual bi-SQUID cells sharing inductances with adjacent cells. We provide extensive computer simulations, analysis and experimental measurements, in which we explore the phase dynamics and linearity of the array voltage response. The non-uniformity in inductances of the bi-SQUIDs produces a pronounced zero-field single anti-peak in the voltage response. The anti-peak linearity and size can be optimized by varying the critical current of the additional junction of each bi-SQUID. The layout implementation of the tight 2D array integration leads to a distinct geometrical ÒdiamondÓ shape formed by the merged dual bi-SQUID cells.  Different size 2D arrays are fabricated using standard HYPRES niobium 4.5 kA/cm$^2$ fabrication process. The measured linearity, power gain, and noise properties will be analyzed for different array sizes and the results will be compared with circuit simulations. We will discuss a design approach for the electrically small magnetic field antenna and low-noise amplifiers with high bandwidth based on these 2D bi-SQUID SQIF arrays. The results from this work will be used to design chips densely and completely covered in bi-SQUIDs that has optimized parameters such as linearity and power gain. 
\end{abstract}


\IEEEpeerreviewmaketitle

\section{Introduction}
\IEEEPARstart{T}{he} quest to increase the linearity of SQUID and SQIF arrays was boosted by the introduction of the bi-SQUID -- a tri-junction dc SQUID with a linear voltage response \cite{KOVbi,Oleg16,Oleg17}. A non-linear inductance of the additional third junction provided the desired linearizing effect. These bi-SQUIDs are now being used in uniform and non-uniform (SQIF) arrays in place of conventional dc SQUIDs with a goal of achieving higher linearity. To date, the most of the design efforts using both SQUIDs and bi-SQUIDs went into the optimization of one-dimensional (1D) serial or parallel arrays and their combinations. This is due to the higher complexity analysis and modeling required for 2D arrays to account mutual coupling of neighboring cells and complex current distribution in arrays.

Arrays of DC bi-SQUIDs with size varying from loop to loop in an unconventional geometric structure are known to exhibit a magnetic flux dependent average voltage response $<V(\Phi_e/\Phi_0)>$, where $\Phi_e$ denotes the external magnetic flux and $\Phi_0$ is the magnetic flux quantum, that has a pronounced single peak with a large voltage swing at zero magnetic flux. The potential high dynamic range and linearity of the ``anti-peak'' voltage response render the array a promising candidate for multiple applications including a detector of absolute strength of external magnetic fields, wide-band low noise amplifiers and magnetic antennas. These arrays are also commonly known as Superconducting Quantum Interference Filters (SQIFs). Since it was theoretically proposed \cite{OHS,HOS} and experimentally demonstrated for the first time \cite{TOH,1211717,OTHN} the SQIF concept is investigated and exploited by a continuously growing number of groups with respect to its basic properties \cite{OHFT,KSOHS,5153080,Oleg1,Oleg2,Oleg3} and its suitability in various fields of application like magnetometry \cite{1211718,SIMH,OCHTFS,COHTFTS,CTOHFTS,PST}, low noise amplifiers of radio frequency (RF) and digital signals \cite{CTOH,5153082,CTOHFTS2,CTOHFTS3,SCO,KCDWVS,Oleg4,Oleg5,Oleg6} and electrically small antennas \cite{electricallysmall,KSKSM}. The unique noise properties, wide bandwidth and linearity of certain configurations of the SQIF array makes it especially attractive for an integration with superconducting wideband Digital-RF receivers being introduced for a variety of military and civilian applications \cite{Oleg7,Oleg8,Oleg9,Oleg10,Oleg11,Oleg12}. These receivers are based on highly linear superconducting analog-to-digital converters \cite{Oleg13,Oleg14,Oleg15} which require equally high linearity of analog RF electronics.

In this paper, we present a new design for the 2D array -- a tightly coupled 2D network of bi-SQUID cells, in which junctions and inductances are shared between adjacent cells. In order to achieve this, we alter the shape (or inductance selection) of the bi-SQUID cells and merge two cells together to form a dual bi-SQUID -- a diamond-shaped bi-SQUID. A diamond shape layout of these cells makes 2D array integration easier and allows for the bi-SQUIDs to be more densely packed. We demonstrate numerically and experimentally that these 2D bi-SQUID SQIF arrays can be optimized to produce a linear anti-peak at the zero magnetic flux. We examine in great detail the voltage response as a function of controlled parameters, including: inductive coupling between loops, number of loops, bias current, and distribution of loop areas. The results from the numerical analysis are validated against experimental results from fabricated designs.

\section{Background: The single SQUID and bi-SQUID}
\begin{figure}
\centering
\includegraphics[width=1.7in]{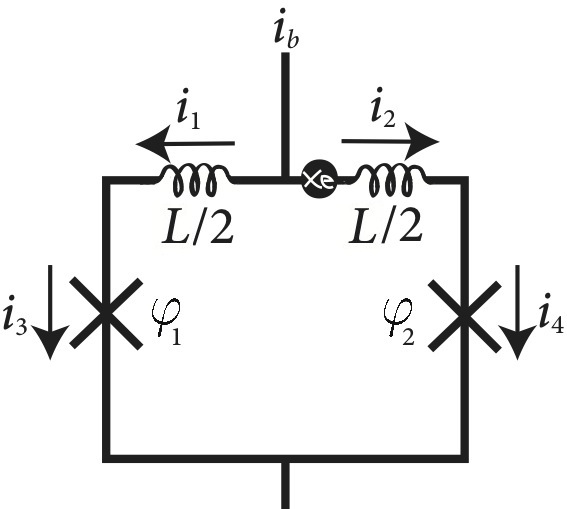}\includegraphics[width=1.7in]{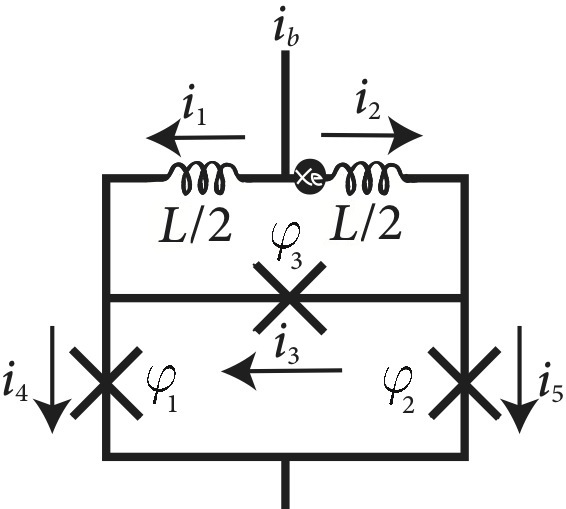}
\vspace{-.3cm}
\caption{Schematic diagram of a single DC SQUID (left) and a single DC bi-SQUID (right) where  ($i_b$, $i_{1}$, $i_{2}$, $i_{3}$, $i_{4}$, $i_{5}$) are the normalized currents,  ($\varphi_{1}$, $\varphi_{2}$, $\varphi_{3}$) represent the phases across the Josephson junctions, $L/2$ is the parameter related to the inductance values and $x_e$ is the point in the equations where the external fields are included.  } 
\label{fig:dcSQU}
\end{figure}

A single DC SQUID has two Josephson junctions arranged in parallel, connected with superconducting material, see the schematic diagram on the left in Fig. \ref{fig:dcSQU}. The equations are derived using Kirchhoff's current law \cite{Kirchhoff}, which is the principle of conservation of electric charge and implies that at any junction in an electrical circuit the sum of currents flowing into that node is equal to the sum of currents flowing out of that node, along with a resistively shunted junction (RSJ) model of the over-damped Josephson junction. 
The Josephson junctions are assumed to be symmetric, in particular having identical critical currents $I_{c1}$ and $i_{c2}$ that we use to normalize all the other currents in our consideration.  The system of equations that models a single SQUID dynamics \cite{memainbiSQUIDs} is

\begin{align} \label{eq:onesys2}
  	&\dot{\varphi}_1 =\ds \frac{i_b}{2}- \frac{1}{L } ( \varphi_1 - \varphi_2 - \varphi_e ) - \sin\varphi_1  \notag\\
   	&\dot{\varphi}_2 =  \ds  \frac{i_b}{2}+\frac{1}{L } ( \varphi_1 - \varphi_2 - \varphi_e ) - \sin\varphi_2,
\end{align}
where $\varphi_1$ and $\varphi_2$ are the phases across each of the Josephson junctions and the dots denote the time differentiation with normalized time $\tau=\omega_c t$, where $t$ is time and $\omega_c=\ds \frac{2e I_c R_N}{\hbar}$. The parameter $R_N$ in $\omega_c$ is the normal state resistance of the Josephson junctions, $I_c$ is the critical current of the Josephson junctions, $e$ is the charge of an electron, and $\hbar$ is the reduced Planck constant.  $i_b=\ds \frac{I_b }{I_c}$ is the normalized bias current, where $I_b$ is the bias current. $L = 2 \pi \ds \frac{l I_c}{\Phi_0}$ is the normalized inductance and $\varphi_e=2\pi a x_e$, where $\ds x_e=\frac{B_{e}}{\Phi_0}$ is the normalized external magnetic flux per unit area and $a$ is the bi-SQUID area. We use the approximate assumption that $a=L$. $\ds \Phi_0 \equiv \ds \frac{h}{2e} \approx 2.07\times10^{-15}$ tesla meter squared is the flux quantum, where $h$ is Plank's constant and $2e$ is the charge on the Cooper pair.

A DC bi-SQUID, which is a SQUID with an additional Josephson junction bisecting the superconducting loop, was introduced as an alternative to traditional SQUIDs and have shown superior linearity in the average voltage response anti-peak feature. The phase equations for the single DC bi-SQUID schematic on the right in Fig. \ref{fig:dcSQU} are derived in a similar way to those of the single DC SQUID. In this case there is a third junction $J_3$ that is related to the first and second junctions through the phases:  $\varphi_1+\varphi_3=\varphi_2$. Using this relationship all the terms that include $\varphi_3$ can be replaced with $\varphi_2-\varphi_1$, thereby reducing the number of phase equations needed to model the system from three to two as shown in \cite{KOVbi}. The governing equations for a single bi-SQUID are
\begin{eqnarray}
\label{eq:biSQUID}
\dot{\varphi}_1 &=&\ds \frac{i_b}{2}-\ds \frac{1}{3L }(\varphi_{1} -\varphi_2-\varphi_e) + \ds \frac{1}{3}i_{c3}\sin(\varphi_2-\varphi_1) \notag\\
&&-\ds \frac{2}{3}\sin\varphi_1-\ds \frac{1}{3}\sin\varphi_2\notag \\
\dot{\varphi} _2 &=& \ds \frac{i_b}{2} +\ds \frac{1}{3L }(\varphi_{1} -\varphi_2-\varphi_e) - \ds \frac{1}{3}i_{c3}\sin(\varphi_2-\varphi_1) \notag\\
&&-\ds \frac{1}{3}\sin\varphi_1-\ds \frac{2}{3}\sin\varphi_2,
\end{eqnarray}
where $i_{c3}=\ds\frac{I_{c3}}{I_c}$, and $I_{c3}$ is the is the critical current of the third junction.  All other parameters are defined as the single DC SQUID.

\begin{figure}[!t]
\centering
\includegraphics[width=3.4in]{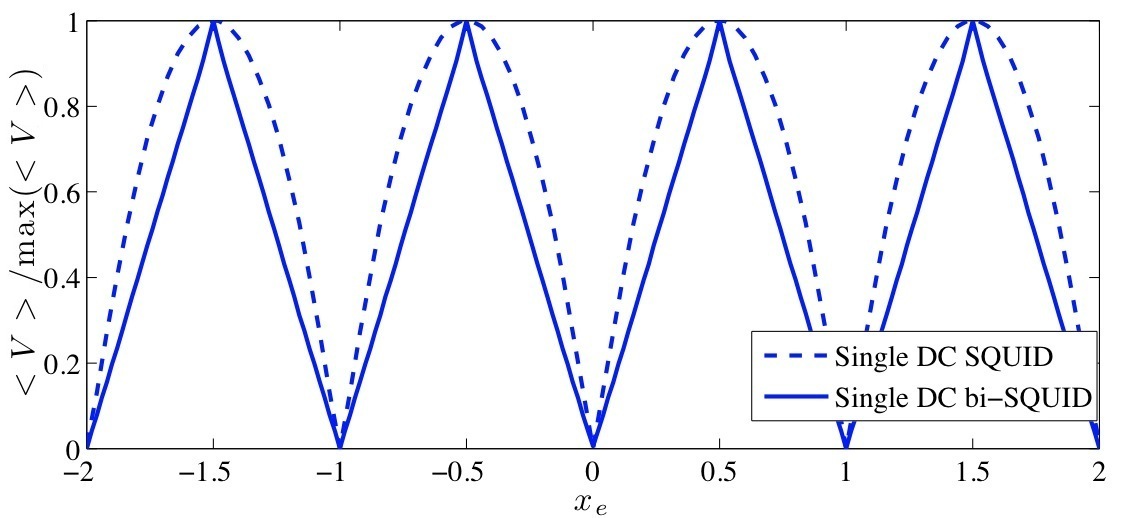}
\vspace{-.3cm}
\caption{Average voltage response of a single DC SQUID compared with that of a single DC bi-SQUID. The computer simulations of the system of equations in Eq. \Ref{eq:onesys2} and Eq. \Ref{eq:biSQUID} were performed with $i_b=2.0$, $L =1.0$ and $i_{c3}=1.0$.} 
\label{fig:AVRsingle}
\end{figure}
The single DC SQUID and DC bi-SQUID are simulated by integrating the systems of equations that model the system dynamics, Eq. \Ref{eq:onesys2} and Eq. \Ref{eq:biSQUID} respectively. After the phases $\varphi_1$ and $\varphi_2$ have been determined the derivatives $\dot{\varphi}_1$ and $\dot{\varphi}_2$ are evaluated and the time-dependant voltage $V(t)=\ds \frac{\dot{\varphi}_1 +\dot{\varphi}_2}{2}$ is calculated. The average voltage, $<V>$, of a SQUID (or bi-SQUID) at a point in $x_e$ is the mean value of the voltage over time, presented in Fig. \ref{fig:AVRsingle}. The average voltage response of the bi-SQUID, with the proper selection of parameters, has a more linear average voltage response than the conventional DC SQUIDs. The higher linearity of the average voltage response increases the utility of the device as a linear amplifier. 

\section{Diamond Shaped bi-SQUID}

\begin{figure}[!t]
\centering
\includegraphics[width=\linewidth]{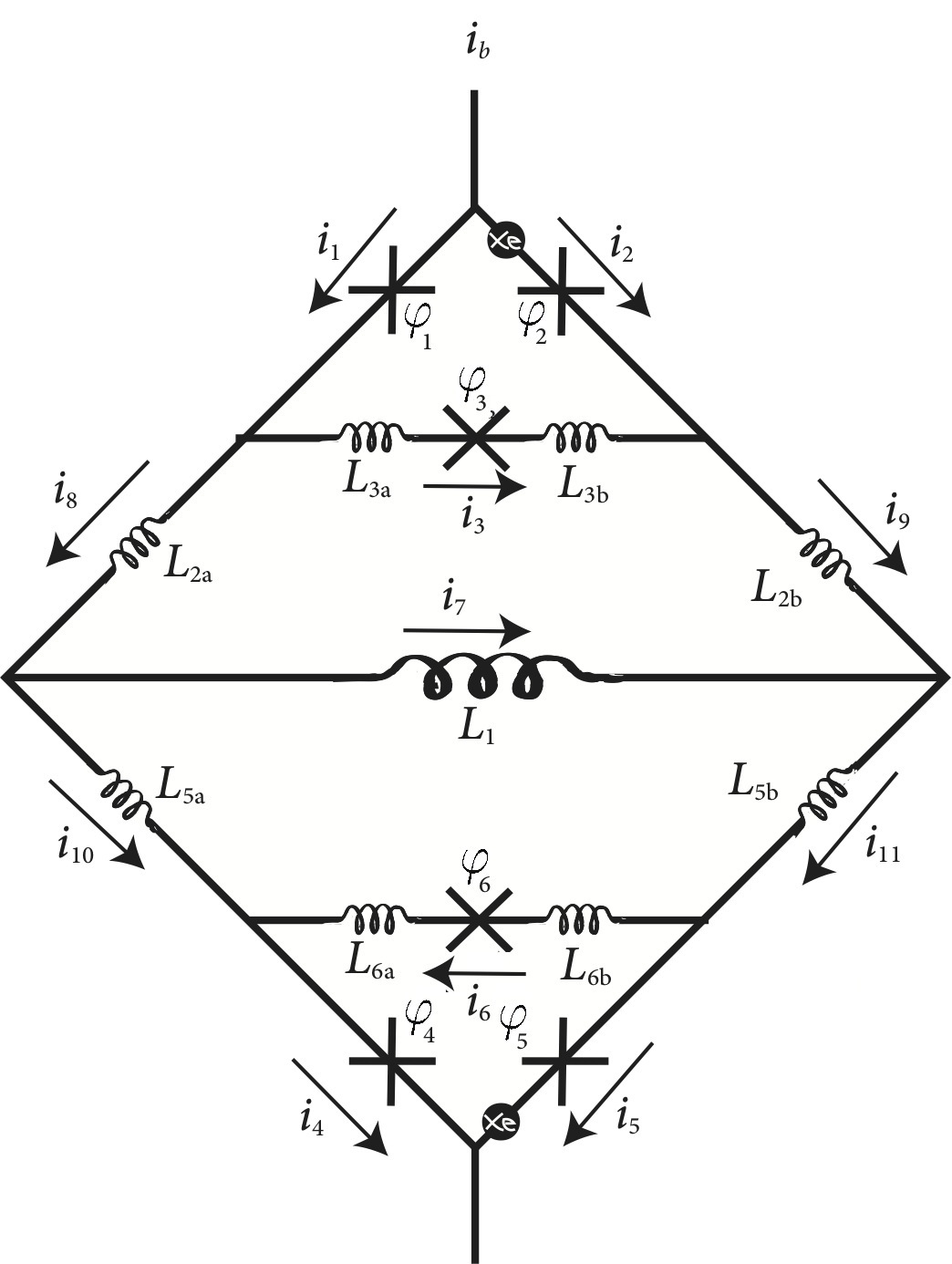}
\vspace{-.3cm}
\caption{Single diamond structure of bi-SQUIDs circuit representation where  ($i_b$, $i_{1}$,...,$i_{11}$) are the normalized currents, ($\varphi_{1}$...$\varphi_{6}$) represent the phases across the Josephson junctions, ($L_{1}$, $L_{2a}$,...,$L_{6a}$, $L_{2b}$,...,$L_{6b}$) are the parameters related to the inductance values and $x_e$ is the point in the equations where the external fields are included. } 
\label{fig:DiaSQUID_s}
\end{figure}

The repeating pattern in the $2$D array is a diamond-shape created by two bi-SQUIDs.  
Fig. \ref{fig:DiaSQUID_s} depicts a circuit of a single diamond structure of bi-SQUIDs where  ($i_b$, $i_{1}$,...,$i_{11}$) represent the normalized currents, ($\varphi_{1}$...$\varphi_{6}$) are the phases across the Josephson junctions and ($L_{1}$, $L_{2a}$,...,$L_{6a}$, $L_{2b}$,...,$L_{6b}$) represent the normalized inductances. To assist with the design of the array, equations that model the system are derived to be used in computer simulations test the effects of different parameter values. Six governing equations are needed to simulate the average voltage response for a single diamond structure of bi-SQUIDs, which contains six Josephson junctions and nine inductors.

Kirchhoff's current law results in the following relations for the currents and phases in the diamond-shaped dual bi-SQUID
\begin{equation} \label{eq:diamcur}
   \begin{array}{lll}
      i_b = i_1 + i_2  & & i_1 = i_3 + i_8  \\ 
      i_2 + i_3 = i_9 & & i_8 = i_7+i_{10}\\ 
      i_9 +i_7 = i_{11} & & i_{10} + i_6= i_4  \\
      i_{11} = i_6 + i_5 & & i_4 +i_5 = i_b \\
      i_1 = \sin \varphi_1 + \dot{\varphi}_1 & & i_2 = \sin \varphi_2 + \dot{\varphi}_2 \\ 
      i_4 = \sin \varphi_4 + \dot{\varphi}_4 & & i_5 = \sin \varphi_5 + \dot{\varphi}_5 \\ 
      i_3 = i_{c3} \sin \varphi_3 + \dot{\varphi}_3 & & i_6 = i_{c6} \sin \varphi_6 + \dot{\varphi}_6, 
   \end{array} 
\end{equation}
where $\varphi_i$ are the phases across of the Josephson junctions, $i=1,...,6$. $I_{ci}$ are the critical currents of each of the junctions, $i=1,...,6$, that satisfy $I_{c1}=I_{c2}=I_{c4}=I_{c5}=I_c$.  $i_{c3} = I_{c3}/I_c$ is the normalized critical current of the third junction and $i_{c6} = I_{c6}/I_c$ is the normalized critical current of the sixth junction. The dot denotes time differentiation with normalized time  $\tau=\omega_ct$. Applying the mesh rule for the phase of the macroscopic wave function of the Cooper pairs to the top half of the diamond gives
\begin{equation}
    \varphi_1 + L_{1} i_7 + L_{2a} i_8  =  2\pi x_e  a_1+ \varphi_2  + L_{2b} i_9,  \notag
\end{equation}
where $\varphi_e=2\pi a_1 x_e$, where $x_e$ is the normalized external magnetic flux and we use the approximate assumption that $a_1=L_{1}+L_{2a}+L_{2b}$. Combining $i_7= i_8 -i_{10} $ and $i_7 = i_{11}-i_9$ such that $i_7= \ds \frac{i_8-i_{10}}{2}+\ds \frac{i_{11}-i_9}{2}$, and then substituting the current relations for $i_7$, $i_{10}$, $i_9$, and $i_8$ yields
\begin{align}\label{eq:diam_main_1}
  &  \left(\ds \frac{L_{1}}{2}+ L_{2a}\right)i_1- \ds \frac{L_{1}}{2}i_4+\ds \frac{L_{1}}{2}i_5+L_{1}i_6 = \varphi_2 -\varphi_1 \quad\quad\notag \\   
  &+ \left(\ds \frac{L_1}{2}+L_{2b}\right)i_2+2\pi x_e  a_1 +\left(L_{1}+L_{2a} + L_{2b}\right)i_3. 
\end{align}
In order to get the first of the six equations that is needed to describe the dynamics of the diamond shape, the bias current relation $i_2=i_b-i_1$ and Josephson junction relations from Eq. \Ref {eq:diamcur} are substituted into Eq. \Ref{eq:diam_main_1}  to become
\begin{align} 
& L_{12} (\dot{\varphi}_1-\dot{\varphi}_3) +\ds \frac{L_{1}}{2}(2\dot{\varphi}_6+\dot{\varphi}_5 -\dot{\varphi}_4)= \left(\ds \frac{L_1}{2}+L_{2b}\right)i_b\notag\\
 &\quad\quad\quad\quad+ \varphi_2- \varphi_1 +2\pi x_e  a_1+ L_{12}\left( i_{c3} \sin \varphi_3 -\sin \varphi_1\right) \notag\\
 &\quad\quad\quad\quad+ \ds \frac{L_{1}}{2}\left( \sin \varphi_4-\sin \varphi_5 -2i_{c6} \sin \varphi_6\right) ,\label{eq:Diamond_first}
\end{align}
where $L_{12}=L_{1} + L_{2a}+L_{2b}$.
To solve for the second of the six equations that governs the dynamics of the single diamond the current relations for the bias current $i_1=i_b-i_2$ and Josephson junctions from Eq. \Ref {eq:diamcur} are substituted into Eq.~\Ref{eq:diam_main_1} to give
\begin{align} 
&\ds \frac{L_{1}}{2}(2\dot{\varphi}_6+\dot{\varphi}_5 -\dot{\varphi}_4)- L_{12} (\dot{\varphi}_2+\dot{\varphi}_3)  =- \left(\ds \frac{L_1}{2}+L_{2a}\right)i_b\notag\\
&\quad\quad\quad\quad+ \varphi_2- \varphi_1 +2\pi x_e  a_1+ L_{12}\left( i_{c3} \sin \varphi_3 +\sin \varphi_2\right) \notag \\
 &\quad\quad\quad\quad+ \ds \frac{L_{1}}{2}\left( \sin \varphi_4-\sin \varphi_5 -2i_{c6} \sin \varphi_6\right). \label{eq:Diamond_second}
\end{align}
The current-phase relation around the upper loop in the single diamond is
\begin{equation} 
\varphi_1+\varphi_3+L_3i_3=\varphi_2, \notag  
\end{equation}
where $L_3=L_{3a}+L_{3b}$. Substituting the Josephson junction relations from Eq. \Ref{eq:diamcur} and reorganizing yields the third of the six equations for the dynamics of the single diamond. 
\begin{align}  \label{eq:Diamond_third}
&L_3\dot{\varphi}_3=\varphi_2-\varphi_1-\varphi_3-L_3i_{c3}\sin \varphi_3.
\end{align}

A similar procedure is used to determine the equations for the three junctions located in the bottom half of the diamond structure and can be combined with the equations from Eqs. \Ref{eq:Diamond_first} - \Ref{eq:Diamond_third} to obtain the full system of equations that governs the phase dynamics of the diamond-shaped bi-SQUID

\begin{align}
	& L_{12} (\dot{\varphi}_1-\dot{\varphi}_3) +\ds \frac{L_{1}}{2}(2\dot{\varphi}_6+\dot{\varphi}_5 -\dot{\varphi}_4)= \left(\ds \frac{L_1}{2}+L_{2b}\right)i_b\notag\\
	&\quad\quad\quad\quad+ \varphi_2- \varphi_1 +2\pi x_e  a_1+ L_{12}\left( i_{c3} \sin \varphi_3 -\sin \varphi_1\right) \notag\\
	&\quad\quad\quad\quad+ \ds \frac{L_{1}}{2}\left( \sin \varphi_4-\sin \varphi_5 -2i_{c6} \sin \varphi_6\right)\notag\\
	&\ds \frac{L_{1}}{2}(2\dot{\varphi}_6+\dot{\varphi}_5 -\dot{\varphi}_4)- L_{12} (\dot{\varphi}_2+\dot{\varphi}_3)  =- \left(\ds \frac{L_1}{2}+L_{2a}\right)i_b\notag\\
	&\quad\quad\quad\quad+ \varphi_2- \varphi_1 +2\pi x_e  a_1+ L_{12}\left( i_{c3} \sin \varphi_3 +\sin \varphi_2\right) \notag \\
	 &\quad\quad\quad\quad+ \ds \frac{L_{1}}{2}\left( \sin \varphi_4-\sin \varphi_5 -2i_{c6} \sin \varphi_6\right) \notag\\
      &L_3\dot{\varphi}_3=\varphi_2-\varphi_1-\varphi_3-L_3i_{c3}\sin \varphi_3  \notag\\
        &L_{15} (\dot{\varphi}_4-\dot{\varphi}_6) + \ds \frac{L_{1}}{2}(2\dot{\varphi}_3 +\dot{\varphi}_2-\dot{\varphi}_1) = \left(L_{5b}+\ds \frac{L_{1}}{2}\right) i_b\notag \\
	&\quad\quad\quad\quad+ \varphi_5- \varphi_4 +2\pi x_e  a_2+ L_{15}\left( i_{c6} \sin \varphi_6 -\sin \varphi_4\right) \notag \\
	&\quad\quad\quad\quad+ \ds \frac{L_{1}}{2}\left( \sin \varphi_1 -\sin \varphi_2-2i_{c3} \sin \varphi_3\right) \notag\\
	& \ds \frac{L_{1}}{2}(2\dot{\varphi}_3 +\dot{\varphi}_2-\dot{\varphi}_1)-L_{15} (\dot{\varphi}_5+\dot{\varphi}_6) =- \left(L_{5b}+\ds \frac{L_{1}}{2}\right) i_b\notag\\
	&\quad\quad\quad\quad+ \varphi_5- \varphi_4 +2\pi x_e  a_2+ L_{15}\left( i_{c6} \sin \varphi_6 +\sin \varphi_5\right)  \notag \\
	&\quad\quad\quad\quad+ \ds \frac{L_{1}}{2}\left( \sin \varphi_1 -\sin \varphi_2-2i_{c3} \sin \varphi_3\right) \notag \\
       &L_6\dot{\varphi}_6=\varphi_5-\varphi_4-\varphi_6-L_6i_{c3}\sin \varphi_3,
\end{align}
where $L_{15}=L_{1} + L_{5a}+L_{5b}$, $L_6=L_{6a}+L_{6b}$ and we use the approximate assumption that $a_2=L_{1}+L_{5a}+L_{5b}$.

\begin{figure}[!t]
	\centering
	\includegraphics[width=3.4in]{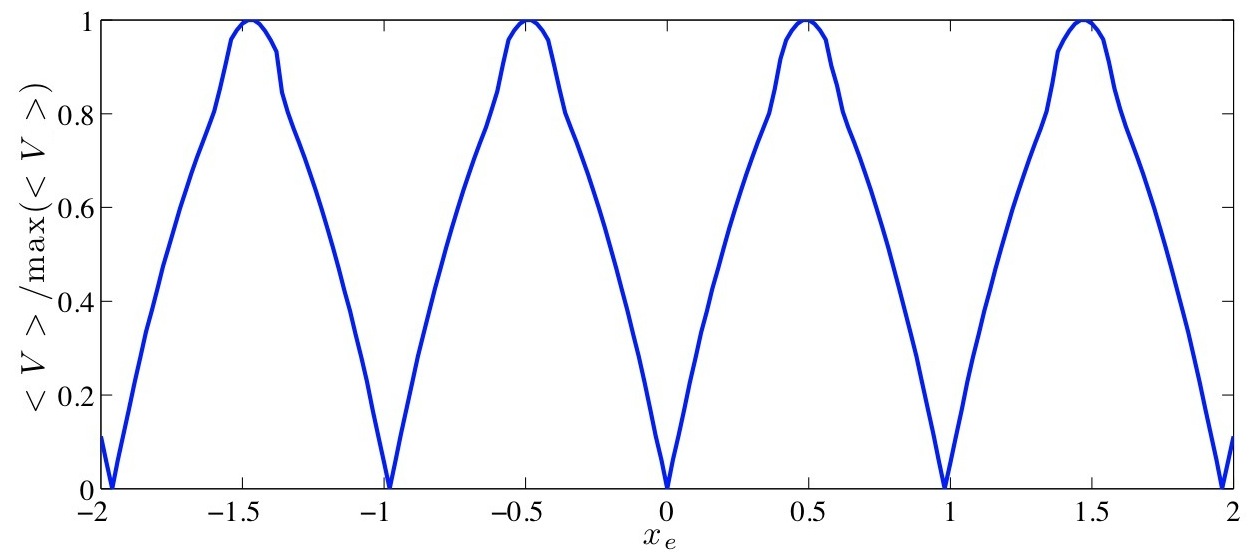}
	\vspace{-.3cm}
	\caption{Calculated average voltage response of a single diamond shaped DC SQUID, performed with $i_b=2.0$, $L_{1}=0.54$, $L_{2a}=L_{2b}=L_{5a}=L_{5b}=0.24$, $L_{3a}=L_{3b}=L_{6a}=L_{6b}=0.3$ and $i_{c3}=0.5$.} 
	\label{fig:AVRtriangle}
\end{figure}

The calculated average voltage response for the diamond shape is shown in Fig. \ref{fig:AVRtriangle}. This average voltage response is more linear than the single SQUID but less so than the single bi-SQUID, shown in Fig. \ref{fig:AVRsingle}. The non-linearities are most likely due to the inclusion of additional inductances in the diamond structure of bi-SQUIDs. The analysis is now extended to the full 2D diamond array. The phase equations that model the 2D diamond arrays are derived in a similar way as the single diamond structure of bi-SQUIDs. These modeling equations total $66$. Both the system of equations and the derivations are not expressed explicitly for brevity.

\section{Implementation of 2D Diamond bi-SQUID Arrays} 
\begin{figure}[!t]
	\centering
	\includegraphics[width=3.4in]{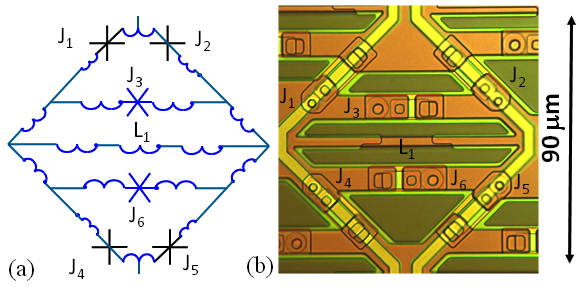}
	\vspace{-.3cm}
	\caption{Diamond-shaped dual bi-SQUID: (a) cell schematic; (b) microphotograph of the fabricated cell within a 2D array. $I_c=0.25$ mA, $R_{\operatorname{sh}}=2.4$ $\Omega$ , $V_c=I_cR_{sh}=600$ $\mu$V. All junctions are critically shunted, $\beta_c=1$.} 
	\label{fig:Oleg1}
\end{figure}

For experimental evaluation of our 2D SQIF arrays of diamond-shaped dual bi-SQUID cells, various size arrays were laid out.  Fig. \ref{fig:Oleg1} shows a schematic and a layout of the diamond-shaped dual bi-SQUID cell designed for a standard HYPRES Nb fabrication process \cite{Oleg18,Oleg19,Oleg20}. $R_{\operatorname{sh}}$ is the shunting resistance and $V_c$ is the critical voltage across each of the junctions. Its electrical circuit was extracted and re-simulated to account for the actual layout parameters. The layout was done using all four available Nb layers: a ground plane layer, two layers for junctions and inductors, and a top layer to implement a flux bias line overlaying bi-SQUID cells (two light gray strips retracing the diamond cell contour).  The ground plane was used only under Nb layers forming bi-SQUID inductors and junctions in order to maintain their low specific inductance. However, ground plane was partially removed from under inductor $L_1$ to increase its value if desired.  The ground plane was also removed from the central area of the bi-SQUID loops to allow an external magnetic field to thread through the cell (dark areas in Fig. \ref{fig:Oleg1}(b)). The total area of the dual-bi-SQUID cell is $\sim$ $1623$ $\mu$m$^2$. While the total area of the cell is kept the same, the values of inductances are varied by their width and ground plane opening size under $L_1$. We designed arrays with normal Gaussian distribution of cell inductances with $\sigma$ $\sim$   $30\%$ and $\sim$ $70\%$ for comparison.

\begin{figure}[!ht]
	\centering
	\includegraphics[width=3.4in]{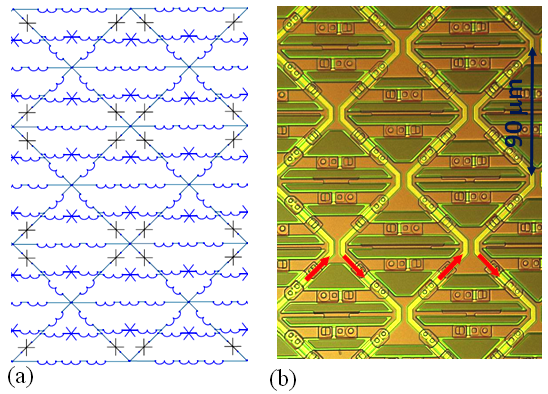}
	\vspace{-.3cm}
	\caption{2D diamond-shaped bi-SQUID array: (a) 2D array schematic; (b) microphotograph of the fabricated 2D array.} 
	\label{fig:Oleg2}
\end{figure}

Fig. \ref{fig:Oleg2} shows a schematic and the layout implementation for a 2D array using diamond dual bi-SQUID cells. It is evident that a diamond shape of our dual bi-SQUID cells leads to their natural 2D arrangement into a diamond checkered pattern by connecting dual bi-SQUID cells by their corners. The space between cells forms a similar array comprising dual bi-SQUIDs with somewhat different ratio of inductors forming bi-SQUID loops.  As a result, each bi-SQUID shares its junctions ($J_1$ and $J_2$) and inductances with neighboring bi-SQUIDs.  Each dual bi-SQUID has a contact to three neighboring cells at each of four corners ($8$ in total).  This 2D array design avoids the use of long parasitic wires, as every component of the array is an essential element of a bi-SQUID indicating to an efficient use of the available area.  

The array dc bias is fed uniformly from the top bi-SQUID row of the array.  Similarly, the array is grounded to the bottom bi-SQUID row.  The inductively coupled flux bias line is overlayed on the top of the array forming loops for each column.  The direction of the dc flux bias control current is shown by small arrows in Fig. \ref{fig:Oleg2}(b).  This line can be used for RF signal input for testing in the low noise amplifier regime, while the output signal is measured at the dc bias current terminal via a bias tee.

\begin{figure}[!t]
	\centering
	\includegraphics[width=3.4in]{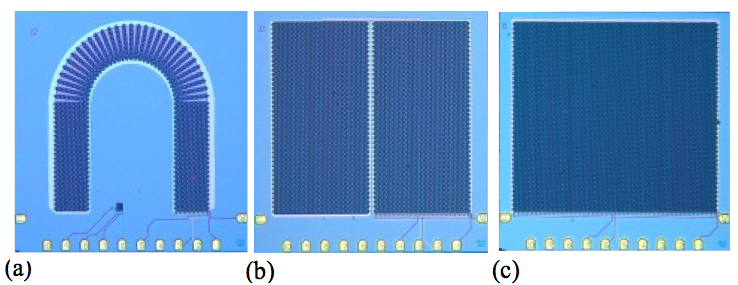}
	\vspace{-.3cm}
	\caption{Microphotographs of $5 \times 5$ mm$^2$ chips of different 2D diamond bi-SQUID SQIF arrays with a $\sigma$ $\sim$ $70\%$ spread: (a) a $1200$ bi-SQUID ($15 \times 80$) array; (b) two serially connected 2D arrays ($2 \times 43 \times 85$) arrays with $7310$ bi-SQUIDs; (c) a single $7820$ bi-SQUID ($92 \times 85$) 2D array. The sizes of these arrays are calculated by assuming each diamond has two bi-SQUIDs.} 
	\label{fig:Oleg3}
\end{figure}

Fig. \ref{fig:Oleg3} shows microphotographs of some examples of the fabricated 2D bi-SQUID SQIF arrays with different size and configuration.  We chose to resistively shunt all junctions in the bi-SQUIDs in the arrays shown. The objective of these designs is to investigate the array dimension dependences. Only $12$ contact pads are used. These are concentrated on one side of the chip in preparation for antenna tests in which RF signal will be irradiated to the chip. To investigate the array characteristics, several test arrays with different sizes and configurations were laid out using $5\times5$ mm$^2$ chips for the fabrication with a $4.5$ kA/cm$^2$ Josephson junction critical current density.

\section{Experimental Evaluation as a Low Noise Amplifier}
Experimental evaluation was first performed in liquid helium using the test setup described in detail in \cite{Oleg21}. Each chip was tested using HYPRES standard cryoprobes with 40 coaxial lines. These cryoprobes can make contact only to the required smaller number (12) of pads (Fig. \ref{fig:Oleg3}) without shorting the array structure by remaining $38$ unused cryoprobe contacts.

\subsection{Experimental Investigation of Noise Properties}

\begin{figure}[!t]
	\centering
	\includegraphics[width=3in]{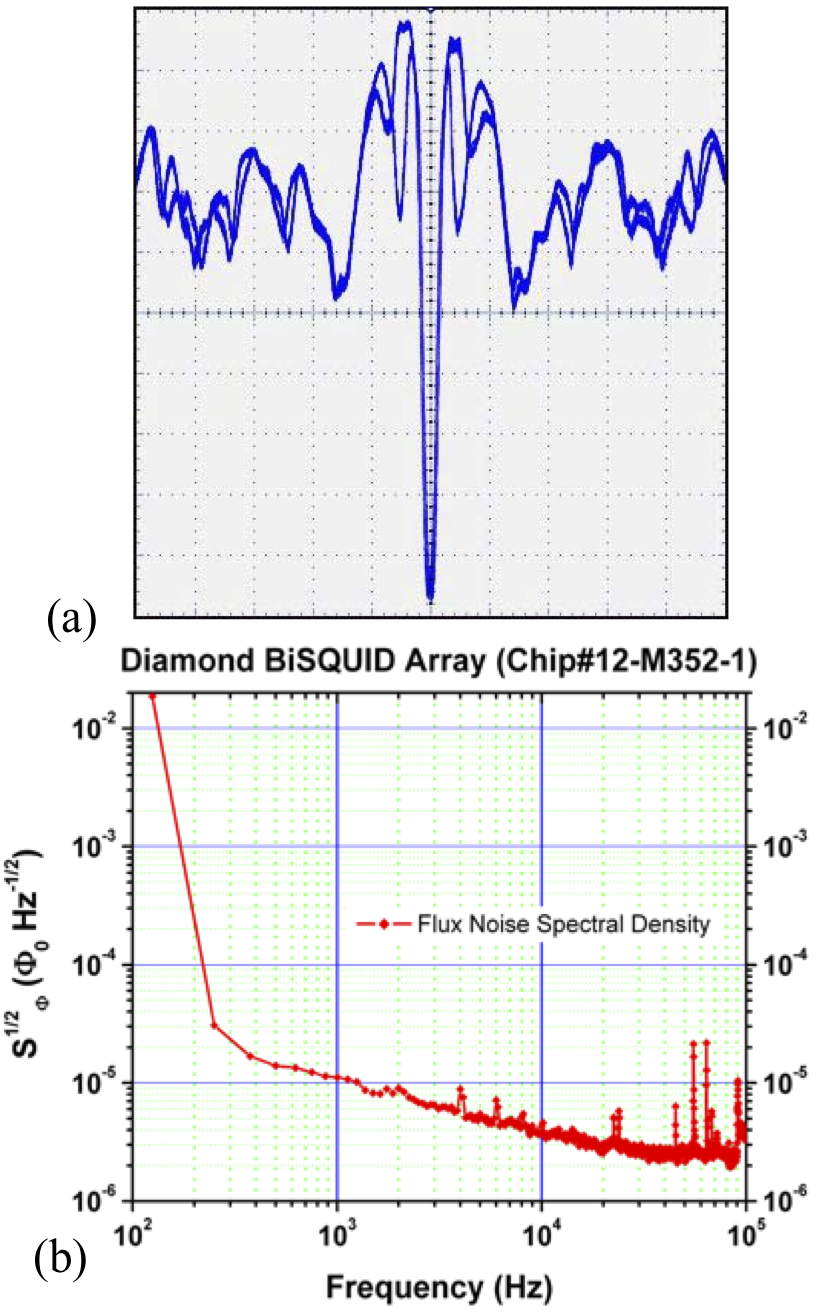}
	\vspace{-.3cm}
	\caption{Measured characteristics of a $15 \times 80$ cell dual bi-SQUID SQIF array of Fig. \ref{fig:Oleg3}(a) with $\sigma$ $\sim$ $70\%$ of inductance spread: (a) flux/voltage characteristic with $2$ mV/div
$0.5$ mA/div, max voltage $\approx 18$ mV, $\Delta$V/$\Delta$I (flux bias) $\approx 170$ V/A.
(b) measured flux noise spectral density. The spikes are attributed to a noisy frequency generator.} 
	\label{fig:Oleg4}
\end{figure}

\begin{figure}[!t]
	\centering
	\includegraphics[width=3in]{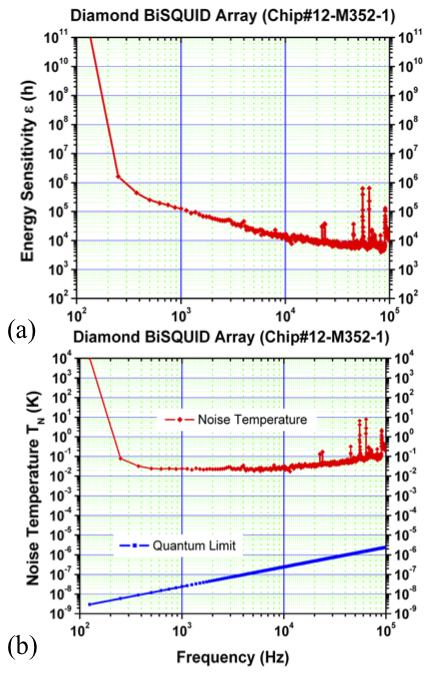}
	\vspace{-.3cm}
	\caption{Noise characteristics of the $1200$ ($15 \times 80$) bi-SQUID SQIF array shown in Fig. \ref{fig:Oleg3}(a): (a) energy sensitivity; (b) noise temperature compared against quantum limit.} 
	\label{fig:Oleg5}
\end{figure}

Fig. \ref{fig:Oleg4}(a) shows the measured flux-to-voltage characteristics of the $15 \times 80$ bi-SQUID array with a $70\%$ inductance spread shown in Fig. \ref{fig:Oleg3}(a). As one can see, it has a well-defined zero-field anti-peak. Fig. \ref{fig:Oleg4}(b) shows the corresponding flux noise measurement at the mid-point of the positive slope of the anti-peak. The flux noise spectral density is reaching to $\sim$ $ 2 \times 10^{-6} \Phi_0 / \sqrt{\operatorname{Hz}}$, which is the expected value for this array. Fig. \ref{fig:Oleg5} shows the corresponding energy sensitivity and noise temperature calculated following \cite{Oleg22,Oleg23}.

The noise energy per unit bandwidth via flux noise in a SQUID is 
\begin{equation}
\varepsilon(f)=\frac{S_\Phi(f)}{2L},
\end{equation}
where $f$ is frequency. The inductance,  $L$, of bi-SQUID is calculated from the measured separately $\Delta I_c$ modulations of the IV curve defined as $L=\frac{\Phi_0}{2\Delta I_c}$.
The noise temperature can be expressed as
\begin{equation}
T_N=\frac{\pi f \varepsilon(f)}{k_B},
\end{equation}
where $k_B$ is Boltzmann's constant. The quantum limit of noise temperature is defined as $T_{QL} = \frac12h f/k_B$ \cite{Oleg24}. The upper frequency ($100$ kHz) for our sensitivity measurements was determined by the available test equipment.  However, the noise temperature has a linear frequency dependence for given device, e. g., for a SQUID RF amplifier, $T_N$ scales as the ratio $\omega_0/V(\Phi)$ \cite{Oleg25}.  So, one can extrapolate the noise data obtained at lower frequencies ($100$ kHz) to higher frequencies and estimate the noise properties up to, e.g., $\sim$ $2$ GHz, which is still much lower than Josephson frequency at the DC bias point.

\subsection{Experimental Results Compared Against Simulations}
\begin{figure}[!t]
	\centering
	\includegraphics[width=3.4in]{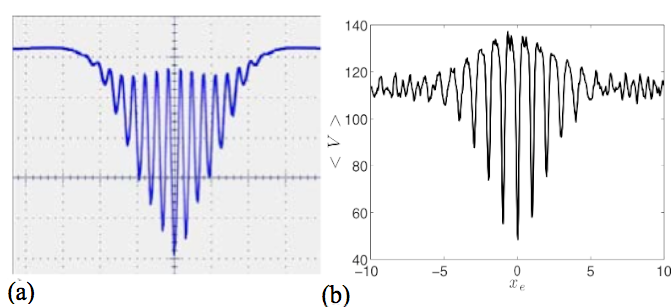}
	\vspace{-.3cm}
	\caption{Comparison of the (a) measured ($5$ mV/div, $10$ mA/div) for a $15\times 40$ dual bi-SQUID array and (b) simulated flux (control current)/voltage characteristic for $\sigma$ $\sim$ $30\%$. The simulations were performed using a $3\times80$ array with $i_b=2.0$ and $i_{c3}=1.0$.} 
	\label{fig:Oleg6}
\end{figure}

\begin{figure}[!t]
	\centering
	\includegraphics[width=3.4in]{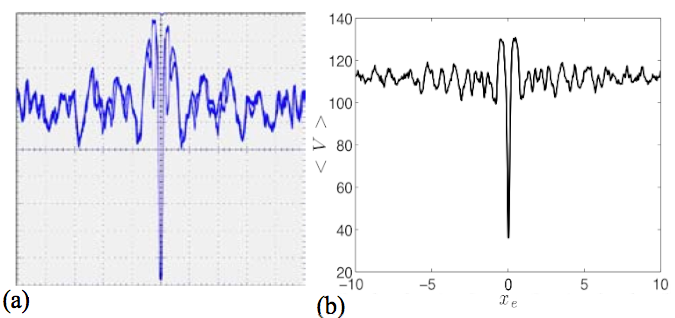}
	\vspace{-.3cm}
	\caption{Comparison of the (a) measured ($2$ mV/div, $1$ mA/div) for a $15\times 80$ dual bi-SQUID array and (b) simulated flux (control current)/voltage characteristic for $\sigma$ $\sim$ $70\%$. The simulations were performed using a $3\times80$ array with $i_b=2.0$ and $i_{c3}=1.0$.} 
	\label{fig:Oleg7}
\end{figure}

For comparison to results of simulation described above, the arrays chips were fabricated with a $30\%$ and $70\%$ Gaussian spread in inductances. Figs. \ref{fig:Oleg6} and \ref{fig:Oleg7} show that increasing the Gaussian spread to $70\%$ makes the average voltage response go from a wave packet-like response to a single anti-peak. The single anti-peak is the form of the average voltage response in which we are interested. The generation of the single anti-peak feature is significant because the anti-peak was obtained with identical size bi-SQUID loops and with varying the inductances. The fabricated arrays that produced the current flux/voltage characteristic in Figs. \ref{fig:Oleg6} and \ref{fig:Oleg7} were $15 \times 40$ and $15 \times 80$ arrays, respectively. While the simulations were performed with arrays of size $3 \times 80$. To simulate larger arrays the code will need to be parallelized.

\begin{figure}[!t]
	\centering
	\includegraphics[width=3.4in]{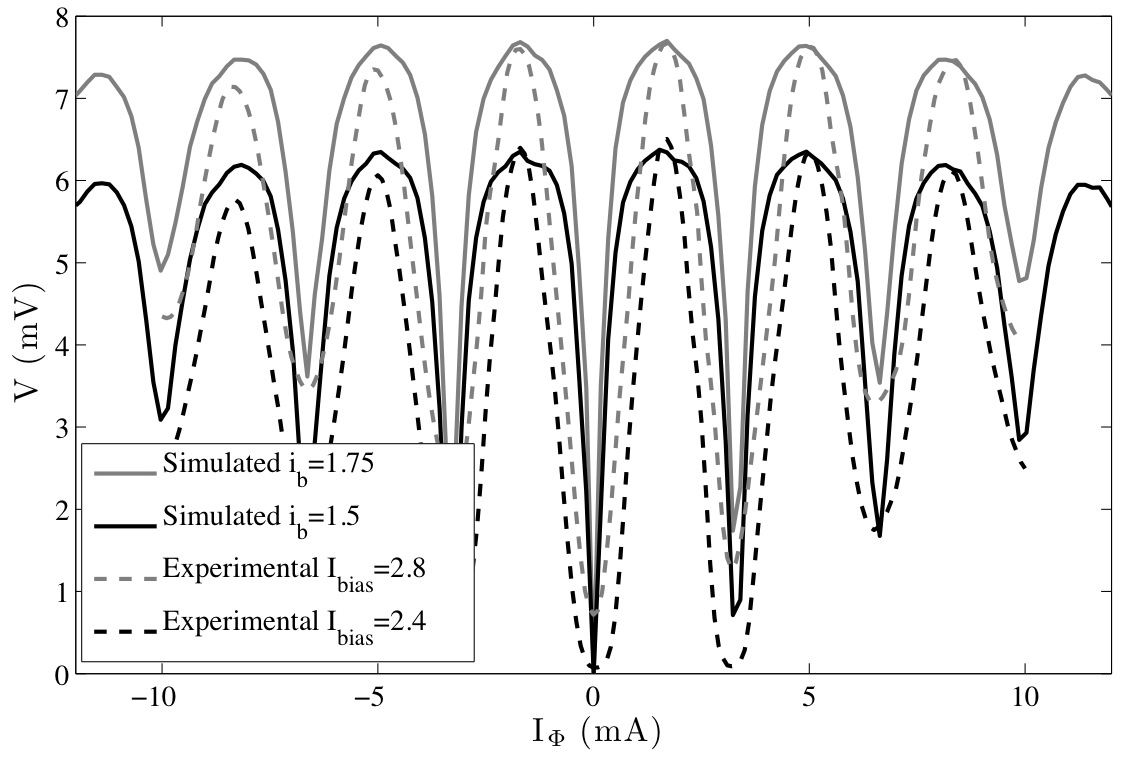}
	\vspace{-.3cm}
	\caption{Comparison of the measured and simulated control current flux/voltage characteristic for a $15\times20$ diamond array with the inductances spread $\sigma$ $\sim$ $30\%$. The measurements of the fabricated array were performed with $I_{bias}=2.4$ and $2.8$. The simulations were performed with $i_b=1.5$ and $1.75$ and $i_{c3}=0.5$.} 
	\label{fig:Compare}
\end{figure}

A comparison of the experimental and simulated current flux/voltage characteristic was performed on an array of size $15\times20$, see Fig. \ref{fig:Compare}. While there are many similarities, the difference in the width of the peaks is most likely due to the array having unshunted third Josephson junctions while the system of equations that model the system are derived using all resistively shunted junctions. 

\subsection{Experimental LNA response}

\begin{figure}[!t]
	\centering
	\includegraphics[width=3.4in]{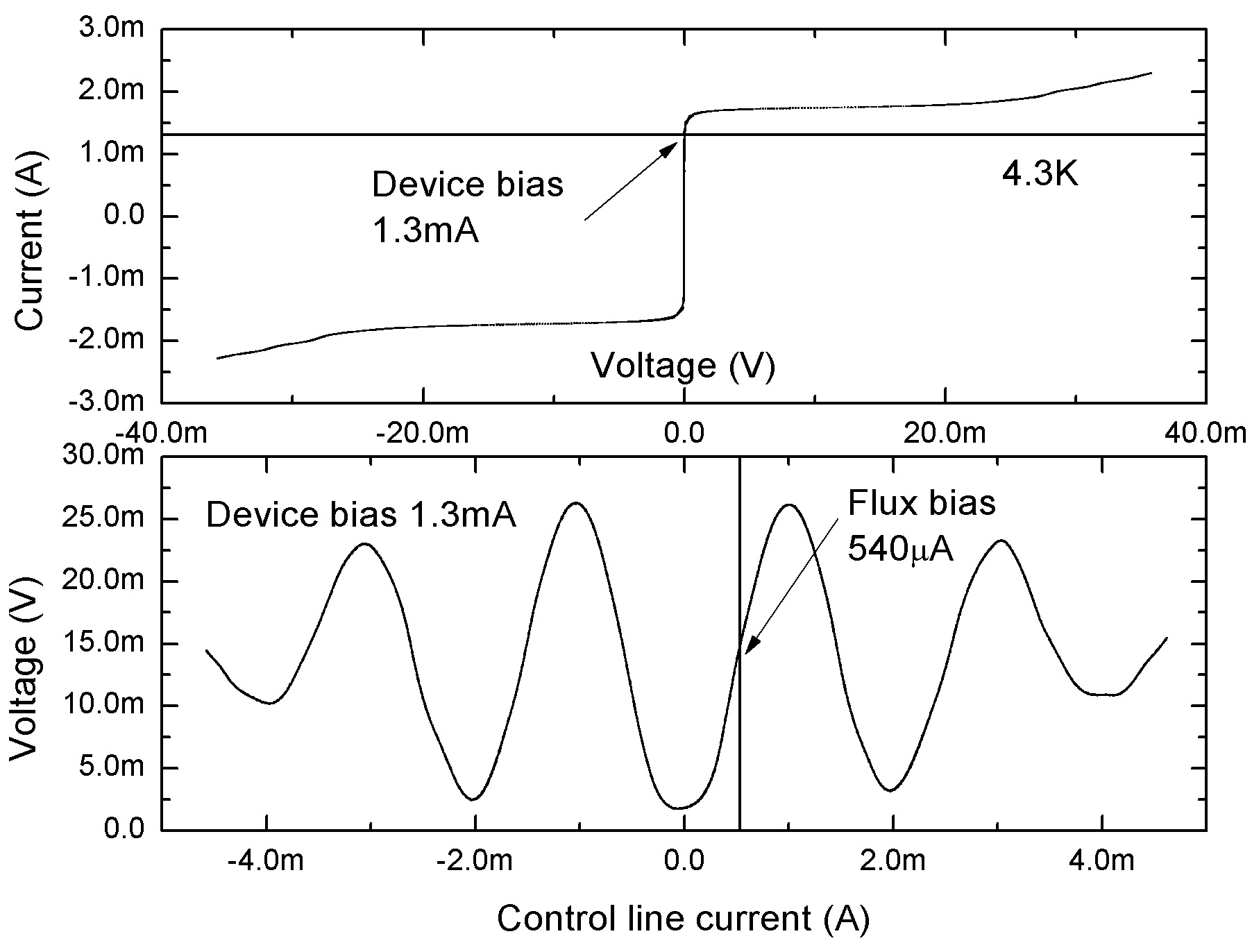}
	\vspace{-.3cm}
	\caption{Visualization of the points of device bias and flux bias for measurements.} 
	\label{fig:Biaspts}
\end{figure}

Measurements of the input power and output power for single tone frequencies were performed $5.3$ MHz, $208$ MHz, and $413$ MHz at the CERF Center of Excellence on a $7 \times 80$ array with a $30\%$ Gaussian spread in inductances and unshunted third junctions, see Table \ref{tab:LNA}. The measurements were performed with the output power normalized to the lowest observable power, which was about $-126.4$ dBm, corresponding to an input power of $-100$ dBm at $208$ MHz, current bias of $1.3$ mA, and flux bias from a control line current of $0.54$ mA, see Fig. \ref{fig:Biaspts}. More measurements will need to be performed to determine how the device is performing as an LNA, however the data will be useful when characterizing devices in the near future (for RF studies).
\begin{table}
\renewcommand{\arraystretch}{1.3}
\caption{SQIF LNA Output to Single RF CW Tones}
\vspace{-5mm}
\begin{center}
\begin{tabular}{cc}
    \hline \hline
    Input Power \@$5.3$ Mhz (dBm)  &  Normalized Power Output (dB) \\
    \hline
    -100   &   --\\
    -80   &   25\\
    -60 & 42.5\\
    \hline
Input Power \@$208$ Mhz (dBm)  &  Normalized Power Output (dB) \\
    \hline
    -100   &   0\\
    -80   &   14\\
    -60 & 34\\
    \hline
    Input Power \@$413$ Mhz (dBm)  &  Normalized Power Output (dB) \\
    \hline
    -100   &   --\\
    -80   &   15\\
    -60 & 34.5\\
    \hline \hline
\end{tabular}\\
\label{tab:LNA}
\end{center}
Table of power input and output values to a single tone at frequencies $5.3$ MHz, $208$ MHz, and $413$ MHz.
\end{table}

\subsection{Antenna Sensitivity Analysis}

We estimated possible antenna sensitivities for a SQIF-based antenna by assuming that the area for the diamond shaped (double) bi-SQUID is $1.62 \times 10^{-9}$ m$^2$. Using that as the effective area, multiplying by $15 \times 80/2$ to account for the number of individual bi-SQUIDs in the array, and taking the flux noise from Fig. \ref{fig:Oleg4}(b) as $2 \times 10^{-6}$ $\Phi_0 /\sqrt{\operatorname{Hz}}$, gives a field sensitivity of $\sim$ $4.25$ fT$/\sqrt{\operatorname{Hz}}$. 

Assuming a scaling as $\sqrt{N}$, the field noise for a $1000 \times 2000$ array would be $0.104$  fT$/\sqrt{\operatorname{Hz}}$ at $100$ kHz. Approximating the physical dimension of the diamond as $\sim$ $71$ $\mu$m $\times \ 71$ $\mu$m 
means that the diamond occupies an area of $\sim$ $5 \times 10^{-9}$ m$^2$ (consistent with the effective area of the diamond shaped bi-SQUID). Thus a $1000 \times 2000$ array would occupy an area of $\sim$ $50$ cm$^2$, since each diamond contains two bi-SQUIDs. This corresponds to a square with $7.1$ cm on a side.

\section{Conclusion}
We have introduced a new design of 2D SQIF arrays using tightly-coupled diamond-shaped dual bi-SQUID cells.  This design allowed us to address the challenge of preserving and enhancing the linearity of individual bi-SQUIDs which tends to be degraded when bi-SQUID cells are connected in parallel. Another advantage of this approach is that the cells can be densely packed onto a chip, resulting in many more bi-SQUIDs in the same area when compared with strings of bi-SQUIDs coupled in series. Our SQIF array design involves bi-SQUIDs with identical areas and varying inductances. We were able to show that a anti-peak feature can be generated with these features. This approach may be applicable for arraying conventional SQUIDs as well.

A system of equations was derived to model the dynamics of the 2D diamond dual bi-SQUID array. The model was very complex and to model large arrays the Matlab code will need to be parallelized. We found a good match between simulations of small arrays and experimental results. The array design has undergone extensive computer simulations, analysis and experimental measurements, in which we explored the voltage response of the 2D arrays.

Experimental measurements were performed on the noise properties of the array design. The results matched the expected values.  More extensive experimentation with larger size arrays of different configurations are required in order to optimize the array design to meet the requirements of wide band low noise amplifiers and electrically small antennas. Ongoing work includes Transverse ElectroMagnetic (TEM) cell measurements to experimentally evaluate the array design as an antenna.

\section*{Acknowledgment}
The authors are grateful to D. Bowling, J. Talvacchio, J. Przybysz for useful discussions, S. Cybart, A. Matlashov, M. Mueck for advice in testing, and HYPRES fabrication team of D. Yohannes, J. Vivalda, R. Hunt, and D. Donnelly for manufacturing the integrated circuits.

\ifCLASSOPTIONcaptionsoff
  \newpage
\fi

\bibliographystyle{IEEEtran}
\bibliography{IEEEabrv,thbib_1230}

\end{document}